\documentclass[letterpaper,10pt]{article} 
\usepackage[utf8]{inputenc}
\usepackage[english]{babel}

\usepackage[letterpaper,top=2cm,bottom=2cm,left=3cm,right=3cm,marginparwidth=1.75cm]{geometry}

\usepackage{romannum}
\usepackage{amsmath}
\usepackage{xcolor}
\usepackage{graphicx}
\usepackage[colorlinks=true, allcolors=blue]{hyperref}
\usepackage{amsmath,amssymb}
\usepackage{epsfig}
\usepackage{caption}
\usepackage{epstopdf}
\usepackage{lipsum}
\usepackage{wrapfig}
\usepackage{lmodern}
\usepackage{cuted}
\usepackage[colorlinks=true, allcolors=blue]{hyperref}
\usepackage{textgreek}
\usepackage[affil-it]{authblk}
\usepackage{hyperref}
\usepackage{xr}
\renewcommand{\textcolor}[2]{#2}

\usepackage[version=3]{mhchem} %
\usepackage{cite}  
\DeclareUnicodeCharacter{2003}{ }
\title{Wavelength-selective nonlinear wavefront control in resonant thin-film lithium niobate metasurfaces}
\author[1]{Madona Mekhael$^*$}
\author[1,2]{Timo Stolt$^*$}
\author[3]{Helena Weigand}
\author[1]{Kiia Arola}
\author[3]{Rachel Grange}
\author[4,5]{Patrice Genevet}
\author[1]{Mikko J. Huttunen}

\affil[1]{Photonics Laboratory, Physics Unit, Tampere University, FI-33014 Tampere, Finland}
\affil[2]{Department of Applied Physics, Aalto University, Aalto FI 00076, Espoo, Finland}
\affil[3]{Optical Nanomaterial Group, Institute for Quantum Electronics, Department of Physics, ETH Zurich, CH-8093 
Zurich, Switzerland}
\affil[4]{Universit{\'e} C\^ote d’Azur, CNRS, CRHEA, 06560 Valbonne, France}
\affil[5]{Physics Department, Colorado School of Mines, Golden, Colorado 80401, USA}
\date{}
\begin{document}
\maketitle
\vspace{-15mm}
\begin{center}
\small\itshape Contact: madona.mekhael@tuni.fi \\
\end{center}
\vspace{-5mm}
\begin{center}
\small\ $^*$These authors contributed equally to this work 
\end{center}

\begin{abstract}

Nonlinear metasurfaces offer compact control over frequency conversion and wavefront shaping. However, existing approaches—often based on geometric phase—lack wavelength selectivity, resulting in static nonlinear responses. Here, we demonstrate a thin-film lithium niobate metasurface that enables spectrally selective shaping of second-harmonic generation through resonance-engineered phase control. The structure consists of two regions with distinct phase responses, realized via spectral tuning of Mie-type resonances. This design enables simultaneous frequency conversion and spatial mode shaping, transforming a Gaussian pump near 1100~nm into a first-order Hermite–Gaussian mode near 550~nm, while maintaining the pump profile. The demonstrated approach offers a pathway toward ultracompact and tunable components for nonlinear holography and related applications.

\vspace{0.3cm}      
\end{abstract}

\section{Introduction}

Nonlinear optical effects are instrumental in developing light sources operating at wavelengths or time scales that are difficult to reach, and have played a key role in emergence of fields such as the generation of nonclassical light~\cite{Weinberg70} and attosecond physics~\cite{Corkum93,Krausz09}. Frequency conversion processes––such as second-harmonic generation (SHG) and sum- or difference-frequency generation––are an important class of nonlinear optical effects with numerous applications, including laser amplifiers~\cite{Shu16,Wise19}, nonlinear microscopy~\cite{Chu16,Bartal21}, quantum light source~\cite{Kumar90,Mook20}, and sensors~\cite{Geissen16,Chandra20,Abajo16}. Traditionally, nonlinear optical effects have been realized in macroscopic nonlinear crystals, which have strict phase-matching requirements for efficient nonlinear responses. Waveguiding media offer a more compact alternative~\cite{Yu14,Gly16,Zhang16,Wan17}, but their performance remains highly dependent on phase matching and is often constrained by limited tunability and narrow operation bandwidths. Moreover, they offer limited flexibility for simultaneous nonlinear conversion and wavefront shaping. To overcome these challenges, metasurfaces have emerged as a promising alternative. These ultrathin optical interfaces are composed of subwavelength structures --- often called meta-atoms --- that allow local control over amplitude~\cite{Staude13,Spin12}, phase~\cite{Over19,Li:19,Gao17}, and polarization~\cite{Li17,Wang21} of light. In the nonlinear regime, metasurfaces benefit from strong local-field enhancement and relaxed phase-matching requirements, enabling highly tunable nonlinear optical processes within a very compact footprint~\cite{GuiLi17}. 

Beyond frequency conversion, nonlinear metasurfaces offer unique capabilities for shaping the spatial profile of the interacting light. In the nonlinear regime, by engineering the phase response of individual meta-atoms, it becomes possible to tailor the wavefront of the generated nonlinear signal, enabling the generation of structured beams, such as higher order Hermite–Gaussian (HG) or Laguerre–Gaussian (LG) modes, or even full nonlinear holograms~\cite{Ren19,Keren-Zur:18,Almeida16,Gao18,Zur16}. 
Metasurfaces can control the phase of the incident light through three key mechanisms: resonant phase shifts, which depend on the phase dispersion near optical resonances \cite{Colom23,Burger24,Mikh23}, geometric (Pancharatnam--Berry) phases, which arise from the rotation of anisotropic elements under circularly polarized excitation~\cite{Chen16,Liu23}, and propagation phases, which result from spatial variations in the optical path length due to changes in the effective refractive index across the metasurface plane~\cite{Chen19}. \textcolor{blue}{While approaches based on geometric phase have enabled impressive demonstrations of nonlinear beam shaping and structured light generation~\cite{Keren-Zur:18, Gao18}, they exhibit only weak wavelength selectivity. Approaches based on propagation phase face a similar limitation~\cite{Yang24}, as the phase profile is determined by the optical path length rather than any spectrally selective feature. In contrast, resonant phase control offers strong wavelength selectivity and field enhancement, making it particularly advantageous for nonlinear optical applications where the spatial mode of the generated harmonic must be controlled at a specific target wavelength.} 

Early realizations of nonlinear metasurfaces utilized metallic nanoparticles, whose nonlinear responses are boosted by strong light confinement at the metal-dielectric interface~\cite{Hooper19,Abir22,Palermo21}. Unfortunately, their performance can be limited by high ohmic losses. Dielectric metasurfaces based on III--V semiconductors with high refractive indices and intrinsic second-order nonlinearities --- such as gallium arsenide~\cite{Sautter19}, aluminum gallium arsenide~\cite{Marino19}, and gallium phosphide~\cite{McL22} --- have recently emerged as low‑loss alternatives, where nonlinear responses can be boosted by Mie-type resonances. However, these materials exhibit absorption in the visible spectral range, limiting their versatility for certain nonlinear applications. In contrast, lithium niobate (LN) overcomes these constraints with its broad transparency window, spanning from the ultraviolet to the mid-infrared. Additionally, LN combines strong second-order nonlinearity, a high electro-optic coefficient, and low optical losses in its thin-film form, making it a widely adopted platform in integrated photonics~\cite{WeisGaylord1985,Loncar21,Kipp23}. \textcolor{blue}{The strong nonlinearity of LN has been leveraged in a variety of resonant nanophotonic geometries, including Mie-resonant metasurfaces~\cite{Setzp20}, quasi-bound-state-in-the-continuum enhanced structures~\cite{Zhang22}, and cavity-based designs~\cite{Li24}, achieving significant enhancement of SHG efficiency. More recently, the rich resonance landscape of LN metasurfaces has been exploited for complex wavefront control, including chiral beam shaping driven by the intrinsic birefringence of LN~\cite{Wang25}.} Despite these advances, it remains challenging to realize LN nanostructures with submicron features, due to the material's resistance to etching. However, recent advances in nanofabrication techniques are enabling the development of LN-based metasurfaces, opening the door to combining its excellent bulk properties with the design flexibility of subwavelength structuring~\cite{Celeb22,Leng24,Grange21,Dionne21,Setzp20}. 

In this work, we combine the nonlinear optical properties of LN with the design flexibility of metasurfaces to demonstrate wavelength-selective spatial beam shaping of the emitted SHG signal. We design and fabricate a metasurface composed of subwavelength LN structures, divided in half with each half featuring different-size nanostructures. These geometry variations induce distinct resonant behaviors, resulting in a spatial phase difference between SHG signal emitted from different sides of the sample. By tailoring the phase profile of the SHG field, we realize multifunctional nonlinear metasurfaces. In particular, we demonstrate combined frequency conversion of pump beam near 1100~nm into SHG beam at 550~nm, together with mode conversion from a Gaussian pump into a first-order Hermite–Gaussian SHG beam. \textcolor{blue}{Crucially, the resonance-phase mechanism exploited here is inherently wavelength-selective --- the HG\textsubscript{01} mode is generated only within a specific spectral window determined by the resonance design, a capability that broadband geometric- and propagation-phase approaches cannot provide.}

\section{Sample design and fabrication}

To identify suitable design parameters, we used Lumerical FDTD to model the optical response of LN nanostructures. The design process began with a parameter sweep over the lattice constant $p$ and the nanoparticle height $h$, aiming to avoid diffraction at the SHG wavelength and ensuring resonance tunability. Based on both fabrication considerations and numerical results, we selected truncated nanopyramids with different dimensions for two regions of the metasurface. In particular, dry etching of LN naturally leads to angled sidewalls, so this geometry ensures good agreement between the designed and fabricated structure. Representative parameter scans are provided in the supplementary material (Figs.~S1 and S2). \textcolor{blue}{Following fabrication, COMSOL Multiphysics simulations were performed using the geometric parameters extracted from SEM characterization. These include transmission spectra and phase responses to validate the resonance-phase mechanism (Fig.~S3), as well as near-field electric field distributions at the resonance wavelengths to characterize the modal nature of the resonances (Fig.~S4).}  

The large metasurface areas of over $0.1~\text{mm}^2$ were fabricated using the combination of soft nanoimprint lithography and top-down etching. This enabled to nanostructure the insulating material stack of LN on quartz without exposing the sample to electron-beam radiation and avoided the need to introduce any metallic charge dissipating layers that can introduce hard to remove impurities. In short, \textcolor{blue}{an x-cut film LN with} of the desired thickness was prepared, followed by the deposition of a silicon nitride hard mask film and a polymer resist soft mask film. The polymer film was molded into the designed nanostructures using thermal imprint lithography and the resulting structures were transferred to the SiN hard mask to ensure sufficient etching selectivity in comparison to the LN film. This was followed by standard LN etching process with inductive plasma etching. \textcolor{blue}{The resulting metasurface was oriented such that its $x$-axis was set along the extraordinary axis of the LN thin film.} More fabrication details can be found in the supplementary information (Fig.~S5). The metasurfaces in this work exemplify well how the approach of utilizing imprint lithography for nanopatterning LN on inert substrates like quartz or sapphire holds significant promise for nonlinear flat photonic devices.

\begin{figure} [t]
    \centering  \includegraphics{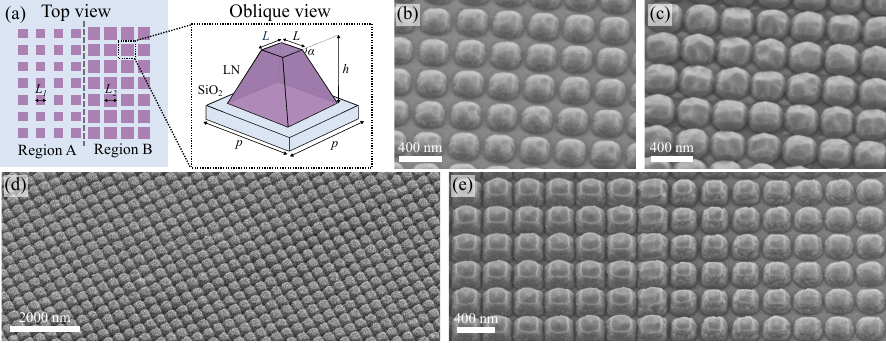}
    \caption{(a) A schematic top view of the metasurface, consisting of two regions (A and B) of truncated LN nanopyramids with a height of $h= 135~\text{nm}$, lattice constant $p=340$~nm an etching angle of $\alpha= 75^{\circ}$, and a varying side length $L$. The inset shows an oblique view of the unit cell. (b, c) Oblique-view SEM images of regions A (\(L_1 = 195~\text{nm}\)) and B (\(L_2 = 250~\text{nm}\)) \textcolor{blue}{taken after removal of the SiN hard mask.} (d, e) Larger-area SEM images from the interface area between regions A and B from a representative metasurface ($L_1=170$\;nm, $L_2=220$\;nm). SEM images were taken at a $30^\circ$ tilt.}
    \label{fig:sample_design}
\end{figure}

As shown in Fig.~\ref{fig:sample_design}(a), our metasurface features a lattice constant $p = 340~\text{nm}$, pyramid height $h = 135~\text{nm}$, and etch angle $\alpha_1 = 75^\circ$ for both regions. Nanoparticles in region A have side length $L_1 = 170~\text{nm}$, while region B is designed with $L_2 = 260~\text{nm}$. Based on our numerical simulations, these parameters induce a phase difference between the two regions of approximately $\pi$ in the visible spectral range, which is sufficient to enable the desired mode shaping. Fig.~\ref{fig:sample_design}(b) and (c) present the Scanning-electron microscopy (SEM) images of the fabricated LN metasurface, showing representative views from both regions of the sample with distinct nanoparticle geometries ($L_1$ and $L_2$). The SEM measurements revealed slight deviations from the design parameters: $L_1 \approx 195$~nm, $L_2 \approx 250$~nm, $h_1 \approx 135$~nm, and $h_2 \approx 120$~nm. \textcolor{blue}{The height difference between the two regions is attributed to the soft nanoimprint lithography process, as discussed in the supplementary material.} The shape of the nanostructures also deviates from ideal truncated pyramids, appearing slightly distorted due to fabrication limitations such as nonuniform etching. These factors collectively impact the precision of the phase control and the fidelity of the generated spatial modes.
 
\section{Linear measurements and phase analysis}
\begin{figure}[t]
\begin{center}
\includegraphics[]{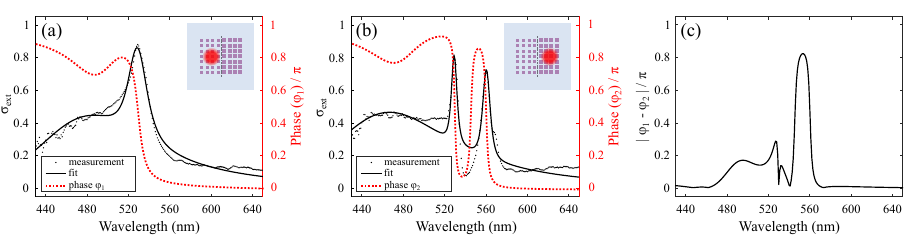}
\end{center}
\caption{(a,b) Measured extinction spectra ($1-T$) from regions A and B of the sample, corresponding to side lengths $L_1 \approx 195$~nm and $L_2 \approx 250$~nm, respectively. The plots show the experimental data (black dots), Lorentzian fits (solid dark lines), and the extracted phase responses (dotted red lines). (c) Phase difference between regions A and B in the visible spectral range.}
\label{fig:T_and_Phases}
\end{figure}

To characterize the metasurface, we measured the transmission spectra $T$ of \textcolor{blue}{$x$-polarized light} from each region of the sample separately. As shown in Fig.~\ref{fig:T_and_Phases}(a), the extinction spectrum $\sigma_\text{ext}=1-T$ from region A reveals a Mie-type resonance near $530~\text{nm}$, while region B exhibits two resonances near $530~\text{nm}$ and $565~\text{nm}$. Due to slight deviations in dimensions and shape between the design and the fabricated nanostructures, the extracted phase response yields a shift close to $\pi$ around a wavelength of 550~nm.  

To extract the phase shifts associated with these resonances, we fitted a Lorentzian function to the measured extinction spectra. \textcolor{blue}{Specifically, the extinction spectrum of region A was fitted using two Lorentzian functions --- one for the resonance near 530~nm and one accounting for a broad background. For region B, three Lorentzian functions were used — two for the resonances near 530~nm and 565~nm, and one for the broad background.} The extinction coefficient $\sigma_\text{ext}$ is linked to the imaginary part of the effective polarizability $\alpha$ through the relation:
\[
\sigma_\text{ext} \propto 4\pi k\, \text{Im}(\alpha),
\]
where $k = \frac{2\pi n}{\lambda}$ is the wavenumber. \textcolor{blue}{ We note that this relation neglects reflectance, which is a valid approximation here. Since no propagating diffraction orders exist ($p = 340$~nm $< \lambda/n \approx 363$~nm at 530~nm), all scattered light is collected in the forward transmission measurement, and $\sigma_\text{ext} = 1 - T$ fully captures the resonant response. Absorption losses are negligible given LN's wide bandgap and the small nanoparticle volume, confirming the validity of the Lorentzian fitting and phase extraction.} Assuming that $\alpha$ follows a complex Lorentzian profile, we first determined $\text{Im}(\alpha)$ from the fitted transmission data. This enabled us to reconstruct the full complex polarizability $\alpha$. The phase shift $\phi$ introduced by the resonances is then obtained from the argument of the complex polarizability:

\[
\alpha = |\alpha| e^{i\varphi},
\]
with $\varphi = \arg(\alpha)$. These phase shifts govern the wavefront manipulation introduced by the metasurface and play a key role in shaping the transmitted light field.

As shown in Fig.~\ref{fig:T_and_Phases}(a) and (b), the retrieved phase shifts span a range from $0$ to $\pi$. For region A, which exhibits a sharp resonance near the wavelength 530~nm, the phase response $\varphi_1$ of this resonance smoothly transitions through $\pi/2$ near the resonance wavelength. In contrast, the phase shift $\varphi_2$ in region B shows a more complex behavior due to the presence of two closely spaced resonances. The interaction between these resonances results in a rapid phase transition around $530~\text{nm}$ followed by a second transition near $565~\text{nm}$. Note that there are no resonances at the pump wavelength range as shown from the transmission spectra at the pump wavelength range in Fig.~S8 in the supplemental material. 

The phase difference between the two regions $\Delta=|\varphi_1-\varphi_2|$ governs the degree of nonlinear wavefront shaping and enables the generation of higher-order spatial modes. As shown in Fig.~\ref{fig:T_and_Phases}(c), at a wavelength of 550~nm, the phase difference reaches a maximum of approximately $0.85\pi$, a condition under which we expect the interacting field to closely resemble the HG\textsubscript{01} mode. 

\section{Nonlinear optical measurements and analysis}

\begin{figure}
    \centering
    \includegraphics{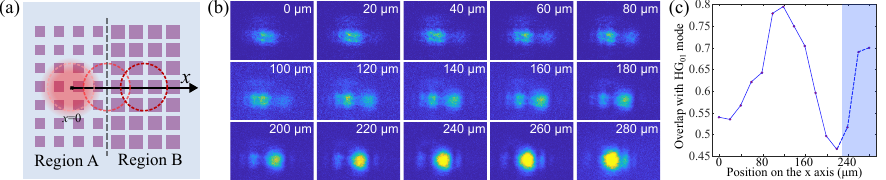}
    \caption{(a) Schematic top view of the sample, indicating the pump beam positions during the scan. The pump starts in region A and is translated along the x-axis in $20~\mu \text{m}$ steps until it reaches region B. (b) Corresponding measured SHG patterns at each pump position, showing the spatial evolution of the SHG mode from a Gaussian profile in region A, gradually transforming into an HG\textsubscript{01}-like mode near the center between the two regions, and reverting back to a Gaussian profile in Region B. (c) Mode-overlap between the measured SHG field and an ideal HG\textsubscript{01} mode, exhibiting a maximum near the center between the two regions. The dashed portion indicates values artificially elevated due to camera saturation.}
    \label{fig:SHG_modes}
\end{figure}

We experimentally verified the generation of HG\textsubscript{01} mode in the SHG signal by pumping the metasurface in the wavelength range of 1000--1300~nm. The experimental setup is described and shown in Fig.~S9 in the Supplementary material. \textcolor{blue}{The pump beam was $x$-polarized, which enables the most efficient intrinsic SHG response via the LN's strongest nonlinear tensor element $d_{33}$.} To investigate how the SHG mode evolves across the sample, we scanned the pump beam along the x-axis from Region A to Region B in steps of approximately $20~\mu \text{m}$, as illustrated in Fig.~\ref{fig:SHG_modes}(a). When the pump is focused entirely within either region, the SHG emission exhibits a fundamental Gaussian (HG\textsubscript{00}) profile. As the pump approaches the boundary between the two regions, the SHG pattern transitions into a higher-order mode consistent with the HG\textsubscript{01} profile. This behavior is quantified in Fig.~\ref{fig:SHG_modes}(c), which shows the overlap between the measured SHG field and an ideal HG\textsubscript{01} mode, revealing a pronounced maximum near the interface. The mode-shaping effect is most clearly observed for a fundamental wavelength around 1100~nm, consistent with the spectral window where the phase difference between the two regions approaches $\pi$. \textcolor{blue}{The broad wavelength range over which SHG is observed is a direct consequence of the relaxed phase-matching requirements in ultrathin metasurfaces. Since the nanoparticle thickness ($\sim$135~nm) is far shorter than the coherence length of LN for the process of SHG, phase mismatch accumulated over the interaction length is negligible across a broad spectral range.}

\begin{figure}
    \centering   
    \includegraphics{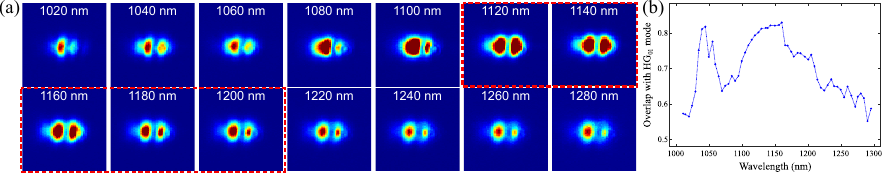}
    \caption{(a) Wavelength-dependent SHG measurements. (b) The overlap with the ideal HG\textsubscript{01} mode as a function of the wavelength. The best mode quality is observed for pump wavelengths between 1100--1200~nm, consistent with the expected phase difference across the metasurface. }
    \label{fig:SHG_wl}
\end{figure}

The observed SHG pattern is shaped by the phase shifts introduced by the distinct nanoparticle geometries of the two metasurface regions. We scanned the pump wavelength and measured the SHG emission at the boundary between the two regions. As shown in Fig.~\ref{fig:SHG_wl}(a), the SHG signal closely resembles the HG\textsubscript{01} within the pump wavelength range of 1100--1200~nm. Fig.~\ref{fig:SHG_wl}(b) shows the corresponding overlap with the ideal HG\textsubscript{01} mode as a function of wavelength, highlighting a maximum near a pump wavelength of 1150~nm. This observation aligns well with theoretical predictions based on the extracted phase difference between the two metasurface regions. \textcolor{blue}{The secondary feature near 1050~nm is attributed to the first resonance of region B near 530~nm, where a partial phase difference between the two regions produces a weaker but observable mode-shaping effect.}

It is worth noting that the measured SHG does not perfectly match the ideal HG\textsubscript{01} mode. This deviation can be attributed to several factors. First, the phase difference between the SHG generated from the two regions is slightly below $\pi$, which is required for ideal mode conversion. While a full $\pi$ phase shift would ideally yield a perfect HG\textsubscript{01} mode, the observed deviation is likely due to the slight shift of the resonances wavelengths from the design, which limits the achievable phase contrast. Second, outside the range where $\Delta\phi \approx \pi$, the SHG output is not a pure Gaussian mode but rather a diffraction pattern induced by the micrometer-scale gap between the two regions (see Fig.~S6(a)). In this sense, the metasurface does not strictly convert a perfect Gaussian beam into an HG\textsubscript{01} mode; rather, it modifies the diffraction pattern to approximate the desired spatial mode. Finally, the SHG intensities generated from the two regions are not perfectly balanced, leading to an asymmetric superposition and residual distortion in the output beam profile. \textcolor{blue}{The estimated peak SHG conversion efficiency is approximately $10^{-10}$ (corresponding to $\sim$20~pW of SHG power) for region A and $3\times10^{-10}$ ($\sim$40~pW) for region B at a pump power of 150~mW, with region B showing higher efficiency consistent with its stronger resonant response.}

\section{Conclusions and outlook}

We have experimentally demonstrated a resonance-based approach for wavelength-selective nonlinear wavefront shaping by introducing a spatially varying phase across a metasurface composed of LN nanostructures. The metasurface features two regions with distinct meta-atom geometries, engineered to produce a distinct resonant phase behavior at the SHG wavelength range of our measurements. By analyzing the linear transmission spectra and extracting the phase response from each region, we identified a spectral window where the difference between phase shifts induced by each region approaches $\pi$ --- a condition that supports the formation of an HG\textsubscript{01}-like mode in the SHG field. Our nonlinear measurements confirmed the formation of the higher-order HG mode near an SHG wavelength of 550~nm, while the pump beam remained unaffected. 

The approach demonstrated here is applicable beyond the specific case of HG\textsubscript{01} generation. Patterning the metasurface in a more complex manner --- for example, radially --- could give rise to LG modes carrying orbital angular momentum~\cite{Fashbender2018}, while other wavefront control functionalities such as lensing, beam steering, and polarization control are also within reach. \textcolor{blue}{In principle, the active device area could be scaled down to tens of micrometers using high-NA focusing conditions, since the Mie-type resonances are dominantly local in nature. However, the hybrid Mie--lattice character of the resonances requires a sufficient number of coherently illuminated unit cells to fully develop, providing a physical justification for using a large excitation beam in the present work.} 

\textcolor{blue}{Extending this approach to nonlocal resonant modes such as quasi-bound states in the continuum (quasi-BICs) is conceptually appealing due to their higher Q-factors and stronger field enhancement~\cite{anthur2020,Ju2023}, which would provide sharper wavelength selectivity and potentially higher SHG efficiency. However, quasi-BICs require long-range periodicity, meaning a spatial boundary between two regions could disrupt the collective mode near the interface, and their high Q-factors make them inherently sensitive to fabrication imperfections. The broader Mie--lattice hybrid resonances demonstrated here therefore offer a more fabrication-tolerant alternative.}

The broad transparency window and strong bulk nonlinearity of LN could be leveraged to extend our method to other nonlinear processes across a wide wavelength range, such as difference-frequency and sum-frequency generation with wavelength-tunable lasers. Active wavelength tunability could further be achieved by exploiting the electro-optical properties of LN to spectrally shift the Mie-type resonances by applying a voltage across the metasurface. Combining this with collective resonances such as quasi-BICs, surface lattice resonances, or guided-mode resonances could further enhance tunability and wavelength selectivity. Overall, the approach demonstrated here provides a pathway toward efficient, tunable, and multifunctional nonlinear optical components for integrated photonics and nanophotonics.

\section*{Acknowledgments}
The authors acknowledge Robert Fickler, Andriy Shevchenko, and Radoslaw Kolkowski for insightful discussions and support, and Ülle-Linda Talts for valuable comments on the manuscript and assistance with fabrication details. T.S. acknowledges financial support from the Jenny and Antti Wihuri Foundation through a doctoral research grant and from the Emil Aaltonen Foundation through a research travel grant. We also acknowledge the Flagship for Photonics Research and Innovation (PREIN) decision number Research Council of Finland/2024/368650 Tampere University. The authors thank the Scientific Center for Optical and Electron Microscopy (ScopeM), the Binning and Rohrer Nanotechnology Center (BRNC), and the FIRST cleanrooms of Eidgenössische Technische Hochschule (ETH) Zurich. This work was supported by the Swiss National Science Foundation SNSF (Consolidator Grant 213713).

\section*{Author declarations}
\subsection*{Conflict of interest}
The authors have no conflicts to disclose.
\subsection*{Author contributions}
M.M. performed numerical simulations, assisted with optical measurements, analyzed the data and wrote the first version of the manuscript. T.S. designed the sample, performed optical measurements, and contributed to manuscript writing. H.W. fabricated the samples and performed SEM measurements. K.A. assisted with transmission measurements, R. G. supervised the sample fabrication. P.G., R.G. and M.H. supervised the project and contributed to manuscript revision. All authors discussed the results and contributed to the final version of the manuscript.

\section*{Data availability}
The data that support the findings of this study are available from the corresponding author upon reasonable request.

\bibliographystyle{unsrt}
\bibliography{references}

@article{Weinberg70,
     author = "David C. Burnham and Donald L. Weinberg",
     title = "Observation of Simultaneity in Parametric Production of Optical Photon Pairs",
     year= "1970",
     journal = {Phys. Rev. Lett.},
     number = "2",
     volume = "25",
}

@article{Corkum93,
     author = "P. B. Corkum",
     title = "Plasma perspective on strong field multiphoton ionization",
     year= "1993",
     journal = {Phys. Rev. Lett.},
     number = "13",
     volume = "71",
}

@article{Krausz09,
     author = "Ferenc Krausz and Misha Ivanov",
     title = "Attosecond physics",
     year= "2009",
     journal = {Rev. Mod. Phys. },
     number = "1",
     volume = "81",
}

@article{Shu16,
     author = "Xuelei Fu and Xiaojie Guo and Chester Shu ",
     title = "Raman-Enhanced Phase-Sensitive Fibre Optical Parametric Amplifier",
     year= "2016",
     journal = {Sci. Rep.},
     number = "20180",
     volume = "6",
    doi = "https://doi.org/10.1038/srep20180"
}

@article{Wise19,
     author = "Pavel Sidorenko and Walter Fu and and Frank Wise",
     title = "Nonlinear ultrafast fiber amplifiers beyond the gain-narrowing limit",
     year= "2019",
     journal = {Optica},
     number = "10",
     pages = "1328--1333",
     volume = "6",
     doi = "https://doi.org/10.1364/OPTICA.6.001328"
}

@article{Chu16,
     author = "Gitanjal Deka and Chi-Kuang Sun and Katsumasa Fujita and Shi-Wei Chu",
     title = "Nonlinear plasmonic imaging techniques and their biological applications",
     year= "2016",
     journal = {Nanophotonics},
     number = "1",
     volume = "6",
     pages = "31--49 ",
     doi = "https://doi.org/10.1515/nanoph-2015-0149"
}

@article{Bartal21,
     author = "Kobi Frischwasser and Kobi Cohen and Jakob Kher-Alden and Shimon Dolev and Shai Tsesses and Guy Bartal",
     title = "Real-time sub-wavelength imaging of surface waves with nonlinear near-field optical microscopy",
     year= "2021",
     journal = {Nat. photon.},
     volume = "15",
     pages = "442--448 ",
     doi = "https://doi.org/10.1038/s41566-021-00782-2"
}

@article{Geissen16,
author = {Martin Mesch and Bernd Metzger and Mario Hentschel and Harald Giessen},
title = {Nonlinear Plasmonic Sensing},
journal = {Nano Lett.},
volume = {16},
number = {5},
pages = {3155--3159},
year = {2016},
doi = {https://doi.org/10.1021/acs.nanolett.6b00478}
}

@article{Chandra20,
author = {Mrigank Singh Verma and Manabendra Chandra},
title = {Nonlinear Plasmonic Sensing for Label-Free and Selective Detection of Mercury at Picomolar Level},
journal = {ACS Sens.},
volume = {5},
number = {3},
pages = {645--649},
year = {2020},
doi = {https://doi.org/10.1021/acssensors.9b02404}
}

@article{Abajo16,
author = {Renwen Yu1 and Joel D. Cox1 and F. Javier Garc\'ia de Abajo},
title = {Nonlinear Plasmonic Sensing with Nanographene},
journal = {Phys. Rev. Lett.},
volume = {117},
pages = {123904},
year = {2016},
doi = {https://doi.org/10.1103/PhysRevLett.117.123904}
}

@article{Kumar90,
author = {Prem Kumar},
title = {Quantum frequency conversion},
journal = {Opt. Lett.},
volume = {15},
number = {24},
pages = {1476--1478},
year = {1990},
doi = {https://doi.org/10.1364/OL.15.001476}
}

@article{Mook20,
author = {Jie Zhao and Chaoxuan Ma and Michael R\"using and Shayan Mookherjea},
title = {High Quality Entangled Photon Pair Generation in Periodically Poled Thin-Film Lithium Niobate Waveguides},
journal = {Phys. Rev. Lett.},
volume = {124},
number = {163603},
year = {2020},
doi = {https://doi.org/10.1103/PhysRevLett.124.163603}
}

@article{Yu14,
author = {Nanfang Yu and Federico Capasso},
title = {Flat optics with designer metasurfaces},
journal = {Nat. Mater.},
volume = {13},
pages = {139--150},
year = {2014},
doi = {https://doi.org/10.1038/nmat3839}
}

@article{Gly16,
title = {Metasurfaces: From microwaves to visible},
journal = {Phys. Rep.},
volume = {634},
pages = {1--72},
year = {2016},
note = {Metasurfaces: From microwaves to visible},
issn = {0370-1573},
author = {Stanislav B. Glybovski and Sergei A. Tretyakov and Pavel A. Belov and Yuri S. Kivshar and Constantin R. Simovski},
doi = {https://doi.org/10.1016/j.physrep.2016.04.004}
}

@article{Zhang16,
author = {Lei Zhang and Shengtao Mei and Kun Huang and Cheng-Wei Qiu},
title = {Advances in Full Control of Electromagnetic Waves with Metasurfaces},
journal = {Adv. Opt. Mater.},
volume = {4},
number = {6},
pages = {818--833},
year = {2016},
doi = {https://doi.org/10.1002/adom.201500690}
}

@article{Wan17,
author = {Weiwei Wan and Jie Gao and Xiaodong Yang},
title = {Metasurface Holograms for Holographic Imaging},
journal = {Adv. Opt. Mater.},
volume = {5},
number = {21},
pages = {1700541},
year = {2017},
doi = {https://doi.org/10.1002/adom.201700541}
}

@article{Hooper19,
     author={David C. Hooper and Christian Kuppe and Danqing Wang and Weijia Wang and Jun Guan and Teri W. Odom and Ventsislav K. Valev},
     year={2019},
     title={Second Harmonic Spectroscopy of Surface Lattice Resonances},
     journal={Nano Letters},
     volume={19},
     number={1},
     pages={165-172},
     isbn={1530-6984},
     doi={10.1021/acs.nanolett.8b03574}
}

@article{Abir22,
author = {Tsafrir Abir and Mai Tal and Tal Ellenbogen},
title = {Second-Harmonic Enhancement from a Nonlinear Plasmonic Metasurface Coupled to an Optical Waveguide},
journal = {Nano Lett.},
volume = {22},
number = {7},
pages = {2712--2717},
year = {2022},
doi = {https://doi.org/10.1021/acs.nanolett.1c04584}
}

@article{Palermo21,
author = {Giovanna Palermo and Massimo Rippa and Ylli Conti and Ambra Vestri and Riccardo Castagna and Giovanna Fusco and Elisabetta Suffredini and Jun Zhou and Joseph Zyss and Antonio De Luca and Lucia Petti},
title = {Plasmonic Metasurfaces Based on Pyramidal Nanoholes for High-Efficiency SERS Biosensing},
journal = {ACS Appl. Mater. Interfaces},
volume = {13},
number = {36},
pages = {43715--43725},
year = {2021},
doi = {https://doi.org/10.1021/acsami.1c12525}
}

@article{WeisGaylord1985,
  author  = {Weis, R.~S. and Gaylord, T.~K.},
  title   = {Lithium Niobate: Summary of Physical Properties and Crystal Structure},
  journal = {Applied Physics A},
  year    = {1985},
  volume  = {37},
  pages   = {191--203},
  doi     = {https://doi.org/10.1007/BF00614817},
}

@article{Sautter19,
author = {J\"{u}rgen D. Sautter and Lei Xu and Andrey E. Miroshnichenko and Mykhaylo Lysevych and Irina Volkovskaya and Daria A. Smirnova and Rocio Camacho-Morales and Khosro Zangeneh Kamali and
Fouad Karouta and Kaushal Vora and Hoe H. Tan and Martti Kauranen and Isabelle Staude and
Chennupati Jagadish and Dragomir N. Neshev and and Mohsen Rahmani},
title = {Tailoring Second-Harmonic Emission from (111)-GaAs Nanoantennas},
journal = {Nano lett.},
volume = {19},
pages = {3905--3911},
year = {2019},
doi = {https://doi.org/10.1021/acs.nanolett.9b01112}
}

@article{Marino19,
author = {Giuseppe Marino and Carlo Gigli and Davide Rocco and Aristide Lema\^{i}tre and Ivan Favero and
Costantino De Angelis and and Giuseppe Leo},
title = {Zero-Order Second Harmonic Generation from AlGaAs-on-Insulator Metasurfaces},
journal = {ACS photonics},
volume = {6},
pages = {1226--1231},
year = {2019},
doi = {https://doi.org/10.1021/acsphotonics.9b00110}
}

@article{McL22,
author = {Blaine McLaughlin and David P. Lake and Matthew Mitchell and Paul E. Barclay},
journal = {J. Opt. Soc. Am. B},
number = {7},
pages = {1853--1860},
publisher = {Optica Publishing Group},
title = {Nonlinear optics in gallium phosphide cavities: simultaneous second and third harmonic generation},
volume = {39},
year = {2022},
doi = {https://doi.org/10.1364/JOSAB.455234}
}

@article{Staude13,
author = {Isabelle Staude and  Andrey E. Miroshnichenko and Manuel Decker and Nche T. Fofang and Sheng Liu and Edward Gonzales and Jason Dominguez and Ting Shan Luk and Dragomir N. Neshev and Igal Brener and Yuri Kivshar},
title = {Tailoring Directional Scattering through Magnetic and Electric Resonances in Subwavelength Silicon Nanodisks},
journal = {ACS Nano},
volume = {7},
number = {9},
pages = {7824--7832},
year = {2013},
}

@article{Spin12,
author = {P. Spinelli and M.A. Verschuuren and A. Polman},
title = {Broadband omnidirectional antireflection coating based on subwavelength surface Mie resonators},
journal = {Nat. Commun.},
volume = {3},
number = {692},
year = {2012},
doi = {https://doi.org/10.1038/ncomms1691}
}

@article{Burger24,
  title = {Poles and zeros in non-Hermitian systems: Application to photonics},
  author = {Felix Binkowski and Fridtjof Betz and R\'emi Colom and Patrice Genevet and Sven Burger},
  journal = {Phys. Rev. B},
  volume = {109},
  issue = {4},
  pages = {045414},
  numpages = {6},
  year = {2024},
  month = {Jan},
  publisher = {American Physical Society},
}

@article{Mikh23,
  title = {Asymmetric phase modulation of light with parity-symmetry broken metasurfaces},
  author = {Elena Mikheeva and R\'{e}mi Colom and Karim Achouri and Adam Overvig and Felix Binkowski and Jean-Yves Duboz and S\'{e}bastien Cueff and Shanhui Fan and Sven Burger and Andrea Al\\{u} and Patrice Genevet},
  journal = {Optica},
  volume = {10},
  issue = {10},
  pages = {1287--1294},
  year = {2023},
  month = {Oct},
  publisher = {Optica Publishing Group},
}

@article{Over19,
author = {Adam C. Overvig and Sajan Shrestha and Stephanie C. Malek and Ming Lu and Aaron Stein and Changxi Zheng and Nanfang Yu},
title = {Dielectric metasurfaces for complete and independent control of the optical amplitude and phase},
journal = {Light Sci. Appl.},
volume = {8},
number = {92},
year = {2019},
doi = {https://doi.org/10.1038/s41377-019-0201-7}
}

@article{Li:19,
author = {Jing Li and Tiesheng Wu and Wenbin Xu and Yumin Liu and Chang Liu and Yu Wang and Zhongyuan Yu and Danfeng Zhu and Li Yu and Han Ye},
journal = {Opt. Express},
number = {16},
pages = {23186--23196},
publisher = {Optica Publishing Group},
title = {Mechanisms of 2$\pi$ phase control in dielectric metasurface and transmission enhancement effect},
volume = {27},
year = {2019},
doi = {https://doi.org/10.1364/OE.27.023186}
}

@article{Gao17,
author = {Song Gao and Wenjing Yue and Chul-Soon Park and Sang-Shin Lee and Eun-Soo Kim and Duk-Yong Choi},
title = {Aluminum Plasmonic Metasurface Enabling a Wavelength-Insensitive Phase Gradient for Linearly Polarized Visible Light},
journal = {ACS Photonics},
volume = {4},
number = {2},
pages = {322--328},
year = {2017},
doi = {https://doi.org/10.1021/acsphotonics.6b00783}
}

@article{Li17,
author = {Tong Li and Xiaobin Hu and Huamin Chen and Chen Zhao and Yun Xu and Xin Wei and Guofeng Song},
journal = {Opt. Express},
number = {20},
pages = {23597--23604},
publisher = {Optica Publishing Group},
title = {Metallic metasurfaces for high efficient polarization conversion control in transmission mode},
volume = {25},
year = {2017},
doi = {https://doi.org/10.1364/OE.25.023597}
}

@article{Wang21,
author = {Shuai Wang and Zi-Lan Deng and Yujie Wang and Qingbin Zhou and Xiaolei Wang and Yaoyu Cao and Bai-Ou Guan and Shumin Xiao and Xiangping Li},
title = {Arbitrary polarization conversion dichroism metasurfaces for all-in-one full Poincar{\'e} sphere polarizers},
journal = {Light Sci. Appl.},
volume = {10},
number = {24},
year = {2021},
doi = {https://doi.org/10.1038/s41377-021-00468-y}
}

@article{GuiLi17,
author = {Guixin Li and Shuang Zhang and Thomas Zentgraf},
title = {Nonlinear photonic metasurfaces},
journal = {Nat. Rev. Mater.},
volume = {2},
number = {17010},
year = {2017},
doi = {https://doi.org/10.1038/natrevmats.2017.10}
}

@article{Liu23,
author = {Bingyi Liu and Ren{\'e} Geromel and Zhaoxian Su and Kai Guo and Yongtian Wang and Zhongyi Guo and Lingling Huang and Thomas Zentgraf},
title = {Nonlinear Dielectric Geometric-Phase Metasurface with Simultaneous Structure and Lattice Symmetry Design},
journal = {ACS Photonics},
volume = {10},
number = {12},
pages = {4357--4366},
year = {2023},
doi =  {https://doi.org/10.1021/acsphotonics.3c01163}
}

@article{Chen16,
author = {Ke Chen and Yijun Feng and Zhongjie Yang and Li Cui and Junming Zhao and Bo Zhu and Tian Jiang},
title = {Geometric phase coded metasurface: from polarization dependent directive electromagnetic wave scattering to diffusion-like scattering},
journal = {Sci. Rep.},
volume = {6},
number = {35968},
year = {2016},
doi = {https://doi.org/10.1038/srep35968}
}

@article{Colom23,
author = {Colom, Rémi and Mikheeva, Elena and Achouri, Karim and Zuniga-Perez, Jesus and Bonod, Nicolas and Martin, Olivier J. F. and Burger, Sven and Genevet, Patrice},
title = {Crossing of the Branch Cut: The Topological Origin of a Universal 2π-Phase Retardation in Non-Hermitian Metasurfaces},
journal = {Laser \& Photonics Reviews},
volume = {17},
number = {6},
pages = {2200976},
doi = {https://doi.org/10.1002/lpor.202200976},
year = {2023}
}

@article{Chen19,
author = {Wei Ting Chen and Alexander Y. Zhu and Jared Sisler and Zameer Bharwani and Federico Capasso },
title = {A broadband achromatic polarization-insensitive metalens consisting of anisotropic nanostructures},
journal = {Nat. Commun.},
volume = {10},
number = {355},
year = {2019},
doi = {https://doi.org/10.1038/s41467-019-08305-y}
}

@article{Loncar21,
     author = "Di Zhu and Linbo Shao and Mengjie Yu and Rebecca Cheng and Boris Desiatov and C. J. Xin and Yaowen Hu and Jeffrey Holzgrafe and Soumya Ghosh and Amirhassan Shams-Ansari and Eric Puma and Neil Sinclair and Christian Reimer and Mian Zhang and Marko Lon\v{c}ar",
     title = "Integrated photonics on thin-film lithium niobate",
     year= "2021",
     journal = "Adv. Opt. Photon.",
     number = "2",
     volume = "13",
     pages = "242--352",
     doi = "https://doi.org/10.1364/AOP.411024"
}

@article{Kipp23,
     author = "Zihan Li and Rui Ning Wang and Grigory Lihachev and Junyin Zhang and Zelin Tan and Mikhail Churaev and Nikolai Kuznetsov and Anat Siddharth and Mohammad J. Bereyhi and Johann Riemensberger and Tobias J. Kippenberg",
     title = "High density lithium niobate photonic integrated circuits",
     year= "2023",
     journal = "Nat. Commun.",
     number = "4856",
     volume = "14",
     doi = "https://doi.org/10.1038/s41467-023-40502-8"
}

@article{Celeb22,
     author = "Anna Fedotova and Luca Carletti and Attilio Zilli and Frank Setzpfandt and Isabelle Staude and Andrea Toma and Marco Finazzi and Costantino De Angelis and Thomas Pertsch and Dragomir N. Neshev and Michele Celebrano",
     title = "Lithium Niobate Meta-Optics",
     year= "2022",
     journal = "ACS Photonics",
     number = "12",
     volume = "9",
     doi = "https://doi.org/10.1021/acsphotonics.2c00835"
}

@article{Leng24,
author = {Runxue Leng and Xingqiao Chen and Ping Liu and Zhihong Zhu and Jianfa Zhang},
journal = {Appl. Opt.},
number = {12},
pages = {3156--3161},
publisher = {Optica Publishing Group},
title = {High Q lithium niobate metasurfaces with transparent electrodes for efficient amplitude and phase modulation},
volume = {63},
year = {2024},
doi = {https://doi.org/10.1364/AO.514979}
}

@article{Grange21,
author = {Helena Weigand and Viola V. Vogler-Neuling and Marc Reig Escal{\'e} and David Pohl and Felix U. Richter and Artemios Karvounis and Flavia Timpu and Rachel Grange},
title = {Enhanced Electro-Optic Modulation in Resonant Metasurfaces of Lithium Niobate},
journal = {ACS Photonics},
volume = {8},
number = {10},
pages = {3004--3009},
year = {2021},
doi = {https://doi.org/10.1021/acsphotonics.1c00935}
}

@article{Dionne21,
    author = {David Barton and Mark Lawrence and Jennifer Dionne},
    title = {Wavefront shaping and modulation with resonant electro-optic phase gradient metasurfaces},
    journal = {Appl. Phys. Lett.},
    volume = {118},
    number = {7},
    pages = {071104},
    year = {2021},
    doi = {https://doi.org/10.1063/5.0039873}
}

@article{Setzp20,
author = {Anna Fedotova and Mohammadreza Younesi and J{\"u}rgen Sautter and Aleksandr Vaskin and Franz J.F L{\"o}chner and Michael Steinert and Reinhard Geiss and Thomas Pertsch and Isabelle Staude and Frank Setzpfandt},
title = {Second-Harmonic Generation in Resonant Nonlinear Metasurfaces Based on Lithium Niobate},
journal = {Nano Lett.},
volume = {20},
number = {12},
pages = {8608--8614},
year = {2020},
doi = {https://doi.org/10.1021/acs.nanolett.0c03290}
}

@article{Ren19,
author = {Haoran Ren and Gauthier Briere and Xinyuan Fang and Peinan Ni and Rajath Sawant and S{\'e}bastien H{\'e}ron and S{\'e}bastien Chenot and St{\'e}phane V{\'e}zian and Benjamin Damilano and Virginie Br{\"a}ndli and Stefan A. Maier and Patrice Genevet},
journal = {Nat. commun.},
number = {2986},
title = {Metasurface orbital angular momentum holography},
volume = {10},
year = {2019},
doi = {https://doi.org/10.1038/s41467-019-11030-1}
}

@article{Keren-Zur:18,
author = {Shay Keren-Zur and Lior Michaeli and Haim Suchowski and Tal Ellenbogen},
journal = {Adv. Opt. Photon.},
number = {1},
pages = {309--353},
title = {Shaping light with nonlinear metasurfaces},
volume = {10},
year = {2018},
doi = {https://doi.org/10.1364/AOP.10.000309}
}

@article{Almeida16,
author = {Euclides Almeida and Ora Bitton and Yehiam Prior},
journal = {Nat. commun.},
number = {12533},
pages = {309--353},
title = {Nonlinear metamaterials for holography},
volume = {7},
year = {2016},
doi = {https://doi.org/10.1038/ncomms12533}
}

@article{Gao18,
author = {Yisheng Gao and Yubin Fan and Yujie Wang and Wenhong Yang and Qinghai Song and Shumin Xiao},
title = {Nonlinear Holographic All-Dielectric Metasurfaces},
journal = {Nano Lett.},
volume = {18},
number = {12},
pages = {8054--8061},
year = {2018},
doi = {https://doi.org/10.1021/acs.nanolett.8b04311}
}

@article{Zur16,
author = {Shay Keren-Zur and Ori Avayu and Lior Michaeli and Tal Ellenbogen},
title = {Nonlinear Beam Shaping with Plasmonic Metasurfaces},
journal = {ACS Photonics},
volume = {3},
number = {1},
pages = {117--123},
year = {2016},
doi = {https://doi.org/10.1021/acsphotonics.5b00528}
}

@article{Fashbender2018,
    author = {Faßbender, Alexander and Babocký, Jiří and Dvořák, Petr and Křápek, Vlastimil and Linden, Stefan},
    title = {Invited Article: Direct phase mapping of broadband Laguerre-Gaussian metasurfaces},
    journal = {APL Photonics},
    volume = {3},
    number = {11},
    pages = {110803},
    year = {2018},
    month = {10},
    issn = {2378-0967},
    doi = {10.1063/1.5049368},
}

@article{Ju2023,
title = {Promoting second-harmonic generation in the LiNbO3 film combined with metasurface using plasmonic quasi bound states in the continuum},
journal = {Photonics and Nanostructures - Fundamentals and Applications},
volume = {57},
pages = {101194},
year = {2023},
issn = {1569-4410},
doi = {https://doi.org/10.1016/j.photonics.2023.101194},
author = {Yao Ju and Wei Zhang and Haoyi Zuo}
}

@article{anthur2020,
  title={Continuous wave second harmonic generation enabled by quasi-bound-states in the continuum on gallium phosphide metasurfaces},
  author={Anthur, Aravind P and Zhang, Haizhong and Paniagua-Dominguez, Ramon and Kalashnikov, Dmitry A and Ha, Son Tung and Ma{\ss}, Tobias WW and Kuznetsov, Arseniy I and Krivitsky, Leonid},
  journal={Nano Letters},
  volume={20},
  number={12},
  pages={8745--8751},
  year={2020},
  publisher={ACS Publications}
}

@article{Yang24,
  author    = {Qian Yang and Xiaona Ye and Haigang Liu and Xianfeng Chen},
  title     = {Highly Efficient and Integrated Nonlinear {Airy} Beams Generation},
  journal   = {Advanced Optical Materials},
  volume    = {12},
  number    = {36},
  pages     = {2401810},
  year      = {2024},
  doi       = {10.1002/adom.202401810}
}

@article{Zhang22,
  author    = {Xiaotian Zhang and Linye He and Xin Gan and Xiaocong Huang 
               and Yixuan Du and Zhenshan Zhai and Zhuang Li and Yuanlin Zheng 
               and Xianfeng Chen and Yangjian Cai and Xianyu Ao},
  title     = {Quasi-bound states in the continuum enhanced second-harmonic 
               generation in thin-film lithium niobate},
  journal   = {Laser \& Photonics Reviews},
  volume    = {16},
  number    = {9},
  pages     = {2200031},
  year      = {2022},
  doi       = {10.1002/lpor.202200031}
}

@article{Li24,
  author    = {Zengya Li and Zhuoran Hu and Xiaona Ye and Zhengyang Mao and Juan Feng and Hao Li and Shijie Liu and Bo Wang and Yuanlin Zheng and Xianfeng Chen},
  title     = {Enhanced Second-Harmonic Generation in Thin-Film Lithium Niobate Circular {Bragg} Nanocavity},
  journal   = {Nano Letters},
  volume    = {24},
  number    = {37},
  pages     = {11676--11682},
  year      = {2024},
  doi       = {10.1021/acs.nanolett.4c03286}
}

@article{Wang25,
  author    = {Bo Wang and Tingyue Zhu and Yunan Liu and Haifang Yang and Ruhao Pan and Junjie Li},
  title     = {Chiral Resonant Modes Induced by Intrinsic Birefringence in Lithium Niobate Metasurfaces},
  journal   = {Physical Review Letters},
  volume    = {134},
  number    = {11},
  pages     = {113802},
  year      = {2025},
  doi       = {10.1103/PhysRevLett.134.113802}
}

\end{document}


\maketitle
\vspace{-15mm}
\begin{center}
\small\itshape Contact: madona.mekhael@tuni.fi
\end{center}
\begin{center}
\small\ $^*$These authors contributed equally to this work 
\end{center}

\section{Numerical simulations} 
\begin{figure}
    \centering
    \includegraphics[width=\linewidth]{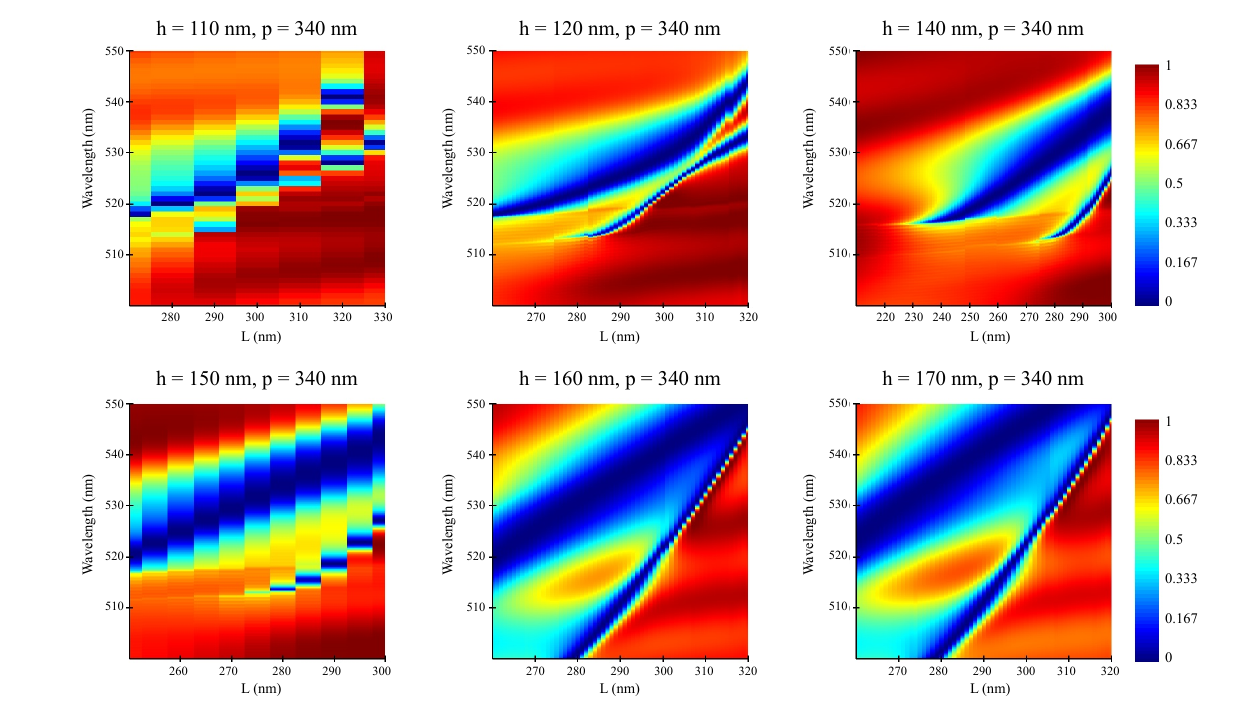}
    \caption{Transmission spectra as a function of the side length $L$ for nanopyramid arrays with $p=340$~nm and a varying height $h$. }
    \label{fig:T_scans}
\end{figure}

\begin{figure}
    \centering
    \includegraphics[width=\linewidth]{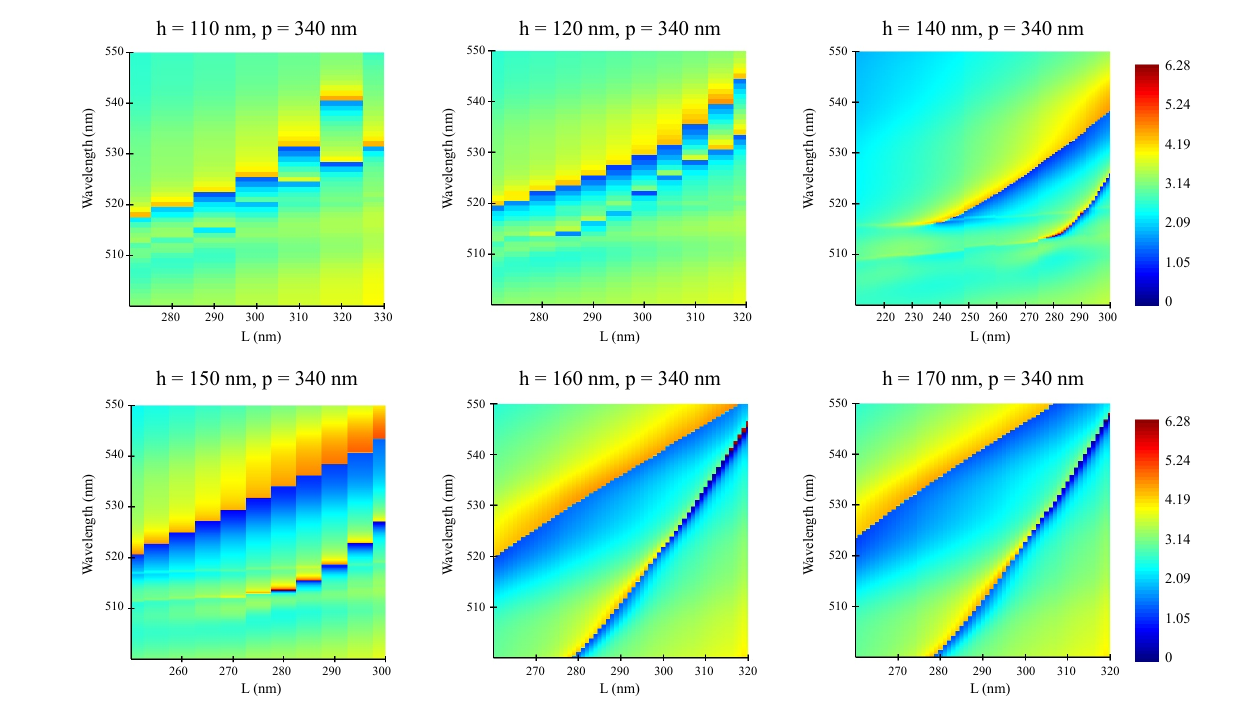}
    \caption{Corresponding phase responses as a function of the side length $L$ for the same structures as in Fig.~\ref{fig:T_scans}. }
    \label{fig:phase_scans}
\end{figure}

Figures~\ref{fig:T_scans} and \ref{fig:phase_scans} show representative numerical scans of the transmission spectra and the corresponding phase responses for a fixed lattice constant $p=340$~nm and a varying particle height between 110--170~nm. For each case, the side length $L$ was swept and the spectral dependence of transmission and phase was extracted. These scans illustrate how the nanoparticle height and lateral dimensions influence the resonance position and the associated phase dispersion, providing the basis for selecting the final design parameters. As observed in Fig.~\ref{fig:T_scans}, increasing the side length of the nanoparticles leads to a redshift of the resonance. At longer $L$ values, a second resonance emerges, which can be tuned by varying the particle height. \textcolor{blue}{This resonance exhibits mixed multipolar character, with contributions from both electric and magnetic dipole modes, consistent with the complex truncated pyramid geometry.} 

\begin{figure}
    \centering
    \includegraphics[width=0.95\linewidth]{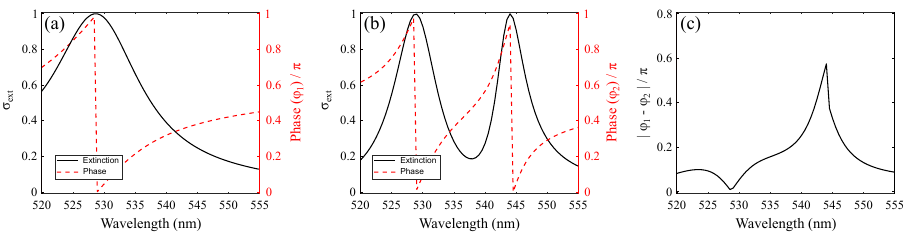}
    \caption{\textcolor{blue}{ Numerical simulations of the transmission spectra and associated phase responses performed using geometric parameters extracted from SEM characterization of the fabricated sample ($L_1 \approx 195$~nm, $h_1 \approx 135$~nm, $L_2 \approx 250$~nm, $h_2 \approx 120$~nm, $p = 340$~nm, etch angle $75^\circ$). (a,b) Simulated extinction spectra and extracted phase responses for regions A and B, respectively. (c) Phase difference between the two regions.} }
    \label{fig:COMSOL}
\end{figure}

\textcolor{blue}{Fig.~\ref{fig:COMSOL} shows numerical simulations performed using geometric parameters extracted from SEM characterization of the fabricated sample, rather than the nominal design parameters. The purpose of these simulations is to validate the resonance-phase mechanism rather than to reproduce the exact resonance wavelengths, which are sensitive to fabrication imperfections and the hybrid Mie--lattice character of the modes. As shown in Fig.~\ref{fig:COMSOL}(a), region A exhibits one resonance near 530~nm within the simulated range, with a smooth phase transition from $\pi$ to zero. Region B, shown in Fig.~\ref{fig:COMSOL}(b), exhibits two resonances within the simulated spectral range near 529~nm and 545~nm, with the phase transitioning through $\pi$ at each resonance. The resulting phase difference, shown in Fig.~\ref{fig:COMSOL}(c), reaches a maximum near the second resonance of region B at 545~nm, confirming the expected resonance-phase behavior that underpins the wavelength-selective wavefront control demonstrated in this work.}

\begin{figure}
    \centering
    \includegraphics[width=0.95\linewidth]{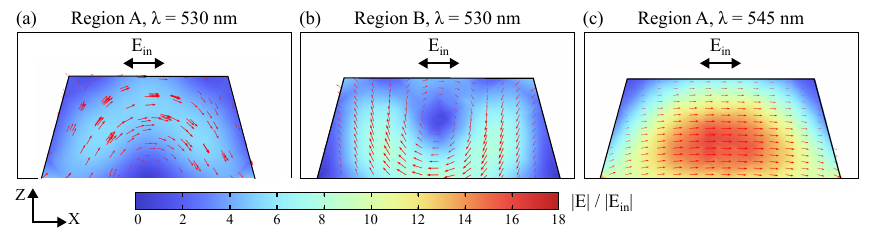}
    \caption{Simulated electric field distributions ($|E|/|E_\text{in}|$) in the $xz$ plane through the center of the nanoparticles at the resonance wavelengths. (a) Region A at 530~nm, (b) Region B at 530~nm, and (c) Region A at 545~nm. Arrows indicate the direction of the local electric field. \textcolor{blue}{The distinct field patterns observed at different resonance wavelengths and between the two regions are consistent with mixed multipolar character, as expected for the truncated nanopyramid geometry. The circulating field pattern in (a) and (b) suggests a magnetic dipole contribution, while the more uniform enhancement in (c) is consistent with electric dipole character.}}
    \label{fig:Efield}
\end{figure}

\textcolor{blue}{Near-field electric field distributions at the resonance wavelengths are shown in Fig.~\ref{fig:Efield}. The distinct field patterns observed at different resonance wavelengths and between the two regions confirm that they support fundamentally different resonant modes, consistent with mixed multipolar character as expected for the truncated pyramid geometry. This difference in modal character between the two regions gives rise to the phase difference exploited for wavelength-selective wavefront control in this work. }

\section{Sample fabrication and characterization}
Fabrication of lithium niobate (LN) nanoparticles on a quartz substrate was carried out using a combination of soft-nanoimprint lithography and reactive ion-etching (RIE). A multilayer stack consisting of quartz, LN (thinned to 155~nm thickness), plasma-enhanced chemical vapor deposited silicon nitride (SiN) (thickness of 310~nm), and a polymethyl methacrylate (PMMA) with a thickness of 160~nm was prepared. The soft-imprint technique was employed, because of challenges to imprint large and isolated areas using direct electron-beam lithography.

\begin{figure} [h]
    \centering
    \includegraphics{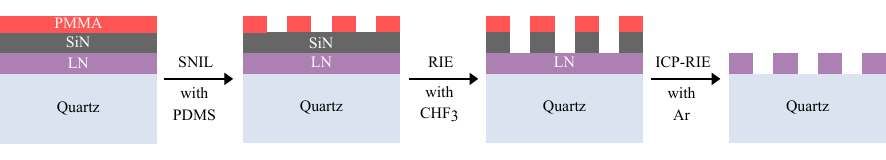}
    \caption{Fabrication steps of the LN nanoparticles on quartz substrate.}
    \label{fig:fab}
\end{figure}

As illustrated in Fig.~\ref{fig:fab}, the nanopatterning via soft nanoimprint lithography began with the preparation of a polydimethylsiloxane (PDMS) mold replicated from a silicon master mold patterned by electron-beam lithography. This mold was then used to transfer the nanoscale patterns onto the top PMMA layer. The transferred pattern was subsequently used as a mask for etching the SiN layer using reactive-ion etching (RIE) with CHF\textsubscript{3}. PMMA alone was unsuitable for pattern transfer due to its low resistance to the intense ion bombardment required for etching the LN layer, which would lead to rapid degradation and loss of nanostructure features. Additionally, the imprint process results in nanostructure size dependent height variation which could be accounted for in the SiN hard mask. Instead, PMMA was used to pattern the SiN layer, which then served as a robust hard mask for etching the LN nanoparticles with RIE and removing the side-product of LN etching in a 70 degrees 40\% potassium hydroxide bath. This process resulted in well-defined LN nanoparticles on the quartz substrate. As the LN film was etched completely to the substrate, the final steps of this nanofabrication require careful handling to not cause detachment of the nanostructures from the substrate.

\textcolor{blue}{The small height difference observed between the two regions in the fabricated sample arises from the soft nanoimprint lithography process. Since the PMMA resist volume per unit cell remains constant during imprinting, meta-atoms with larger lateral dimensions (region B) result in a thinner PMMA resist layer. This leads to a slightly thinner SiN hard mask for region B and consequently a slightly reduced etch depth during LN etching, resulting in shorter nanoparticles in region B compared to region A.} As the ICP-RIE etching consumed nearly the full thickness of the SiN mask.  Future devices produced using this method can avoid this artefact by increasing the SiN hard mask thickness.

\begin{figure}
    \centering
    \includegraphics{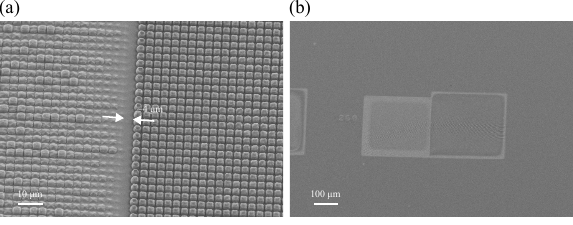}
    \caption{(a) SEM image showing the two regions of the measured sample with a fabrication-induced gap of approximately $4~\mu$m between them, which acts as a diffraction wire distorting the spatial mode of the SHG beam. (b) Large-area SEM image of the metasurface with a total area of $0.15~\text{mm}^2$. SEM images were taken at a $30^\circ$ tilt. }
    \label{fig:SEM}
\end{figure}

Fig.~\ref{fig:SEM}(a) presents a Scanning-electron microscopy (SEM) image showing the two regions of the sample separated by a narrow gap of approximately $4~\mu$m, which was introduced during fabrication. This gap, a result of the imprint process where the resist did not fully fill the mold at geometry transitions, acts as a diffraction wire. It explains why the SHG beam profile deviates from a Gaussian shape at wavelengths far from the $\pi$---particularly when the phase difference between the two regions approaches zero. It is worth noting that the gap is less apparent in some designs with different side lengths values. Fig.~\ref{fig:SEM}(b) shows a large-area view of the metasurface, covering a total area of $0.15~\text{mm}^2$. To our knowledge, such a large continuous region of patterned LN metasurface was not previously reported.

\section{Experimental setup}
\begin{figure}
    \centering
    \includegraphics{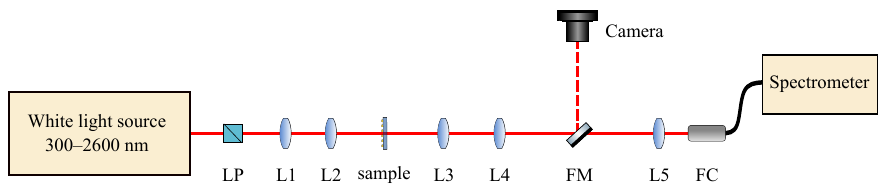}
    \caption{A schematic for the setup used to measure the transmission spectra from the sample.}
    \label{fig:T_setup}
\end{figure}

\begin{figure}
    \centering
    \includegraphics[width=0.5\linewidth]{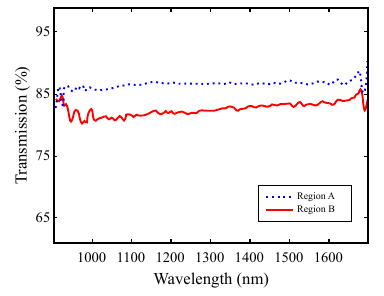}
    \caption{Transmission spectra from the two regions of the sample at the pump wavelength range.}
    \label{fig:pump_T}
\end{figure}

To measure the transmission spectra from the sample, we used the setup illustrated in Fig.~\ref{fig:T_setup}. A white-light source covering the range 300--2600~nm provided illumination, and a broadband linear polarizer (LP) was used to control the polarization. The sample was positioned in the Fourier plane of a $4f$ system consisting of lenses L1--L4 with focal lengths $f_1=40~\text{mm}, f_2=18~\text{mm}, f_3=19.5~\text{mm},$ and $f_4=75~\text{mm}$. A flip mirror (FM) directed the beam to a camera for sample imaging, while lens L5 coupled the transmitted light into a spectrometer via a fiber coupler (FC). The transmission spectra were recorded with an AvaSpec-ULS-RS-TEC spectrometer (Avantes). Fig.~\ref{fig:pump_T} shows the transmission spectra from Regions A and B of the sample, indicating that no resonances are present within the pump wavelength range. Consequently, the pump beam remains unshaped, and the observed variations in the SHG mode arise from resonances occurring at the SHG wavelength.

\begin{figure}
    \centering
    \includegraphics{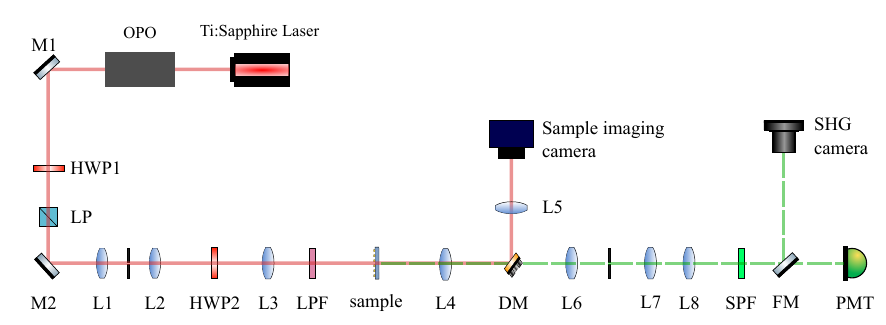}
    \caption{A schematic for the optical setup used to measure the SHG from the sample and image its spatial modes.}
    \label{fig:setup}
\end{figure}

Figure~\ref{fig:setup} shows the experimental setup used to measure the nonlinear signal from the sample. An optical parametric oscillator (OPO) pumped by a femtosecond laser (Chameleon Vision \MakeUppercase{\romannumeral 2}) served as the tunable light source, providing pulses of 220~fs duration within the 1000–1300~nm wavelength range at a repetition rate of 82~MHz. The power was controlled and stabilized during wavelength scanning using a motorized half-wave plate (HWP1) in combination with a polarizer. To ensure high beam quality, the laser beam was spatially filtered using a pair of lenses with a pinhole positioned between them, which effectively removed higher-order spatial modes before the beam interacted with the sample.

\begin{figure}
    \centering
    \includegraphics[width=0.5\linewidth]{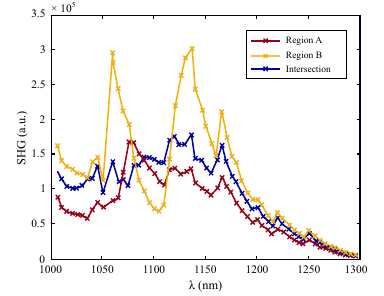}
    \caption{Measured SHG spectra from different regions of the sample. Measurements show agreement with the measured transmission spectra.}
    \label{fig:PMT_measurement}
\end{figure}

A lens L3 ($f=500$~mm) lens was used to focus the pump beam onto the sample, with an additional half-wave plate (HWP2) to control its polarization. To block any unwanted light at the SHG wavelength range, a long-pass filter was placed before the sample. After interacting with the sample, the emitted SHG signal was separated from the fundamental beam using a dichroic mirror, which reflected the pump beam while transmitting the SHG. The reflected pump was directed through a $4f$ imaging system composed of lenses L4 and L5 to image the sample. Meanwhile, the transmitted SHG signal was directed toward the SHG camera for spatial mode characterization. Two consecutive $4f$ imaging systems, with a pinhole placed at the intermediate focal plane, were used to filter and image the SHG mode immediately after the sample. A short-pass filter was used in front of the camera to block any remaining pump light and ensure that only the SHG signal was imaged. A flip mirror (FM) was used to switch between the photomultiplier tube (PMT) to measure the strength of the signal and the SHG camera to image the spatial profile of the SHG beam. Fig.~\ref{fig:PMT_measurement} shows the SHG signal measured using the PMT at the left (region A), right (region B), and the center of the sample between the two regions. The results show good agreement with the design as well as the measured transmission spectra. 

\begin{figure}
    \centering
    \includegraphics[width=0.55\linewidth]{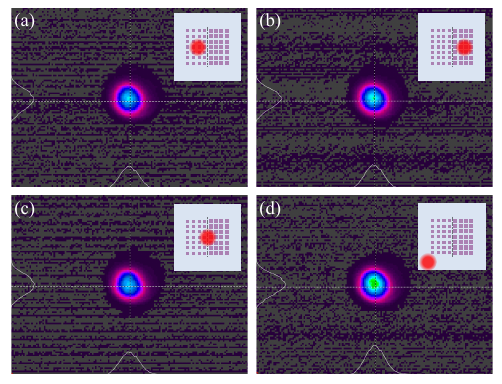}
    \caption{Pump beam spatial profile after interacting with (a) region A. (b) region B. (c) the center between the two regions. (d) the glass substrate. The pump beam retained its Gaussian profile in all cases confirming that spatial mode shaping occurs only for the SHG signal.}
    \label{fig:pump_beam}
\end{figure}

To demonstrate that our device selectively shapes only the SHG signal without affecting the pump beam, we imaged the spatial profile of the pump after transmission through different parts of the sample using a beam profiler (Pyrocam~\MakeUppercase{\romannum{3}} HR from Spiricon)(Fig.~\ref{fig:pump_beam}). The measurements were performed after the beam interacted either with Region A, Region B, the center between regions A and B, or a bare glass substrate. In all cases, the transmitted pump retained its Gaussian profile, confirming that the metasurface does not modify the spatial shape of the fundamental beam. This result validated that the mode shaping occurs exclusively at the SHG wavelength, as intended, and not due to any reshaping of the input beam.

